\documentclass[a4paper]{PoS}
\usepackage{xspace}
\usepackage{xcolor}
\usepackage{wrapfig}

\def\fig#1{fig.~\ref{fig:#1}}
\def\ppS{$\sqrt{s}~=~7~\textrm{TeV}$\xspace}
\def\pT{\ensuremath{p_{\textrm{\tiny T}}}\xspace}
\def\S#1{$\sqrt{\textrm{s}_{\textrm{\tiny NN}}}~=~#1~\textrm{TeV}$\xspace}
\def\dNdeta{$\textrm{d}N_{\textrm{\tiny ch}}/\textrm{d}\eta$\xspace}
\title{New results related to QGP-like effects in small systems with ALICE}

\ShortTitle{New results related to QGP in small systems with ALICE}

\author{\speaker{Vytautas Vislavicius} for the ALICE collaboration\\
        Lund University\\
        E-mail: \email{vytautas.vislavicius@cern.ch}}

\abstract{Results on the production of $\pi^{\pm}$, $\textrm{K}^{\pm}$, $\textrm{p}(\bar{\textrm{p}})$, $\Lambda(\bar{\Lambda})$, $\Xi^{-} \left(\bar{\Xi}^{+}\right)$ and $\Omega^{-} \left(\bar{\Omega}^{+}\right)$ at midrapidity (${|y|<0.5}$) as a function of multiplicity in \ppS pp collisions are reported. Transverse momentum distributions and integrated yields are compared to expectations from statistical hadronization models along with results from different colliding systems and center-of-mass energies. The evolution of spectral shapes with multiplicity show similar patterns to those seen in p-Pb and Pb-Pb collisions. The \pT-integrated baryon yields relative to pions exhibit a significant strangeness-related enhancement in both pp and p-Pb collisions.}

\FullConference{Fourth Annual Large Hadron Collider Physics\\
		13-18 June 2016\\
		Lund, Sweden}

\begin{document}
\section{Introduction}
A strongly coupled deconfined medium of quarks and gluons, the Quark Gluon Plasma (QGP), is created in ultra-relativistic heavy-ion collisions at the Large Hadron Collider. There is strong evidence that this medium is in local thermodynamic equilibrium and is subject to a collective behaviour of the emitted particles, including a common velocity boost in the radial direction (radial flow). Recently, similar patterns have been observed in p-Pb collisions at \S{5.02}~\cite{pPb_PbPb_pOverpi}, which poses a natural question whether any signs of collective behaviour can also be found in even smaller systems, as for instance pp. 

While the double-ridge structures from two-particle correlation studies observed in pp, p-Pb and Pb-Pb have been reported in other publications~\cite{pPb_Ridge,pp_Ridge}, the following proceedings will focus on hadrochemistry and transverse momentum (\pT) spectra of identified particles. A Large Ion Collider Experiment (ALICE)~\cite{ALICE_Det} is particularly suitable for such studies due to its excellent particle identification (PID) capabilities over three orders of magnitude in transverse momentum, from 100~MeV/{\it c} to 20~GeV/{\it c}.

\section{Analysis}

The analysis presented in the following proceedings is performed on a sample of $2.8\cdot 10^{8}$ minimum bias pp collision events at \ppS recorded with ALICE during the 2010 LHC data taking. Detectors relevant to this analysis included the Inner Tracking System (ITS), the Time Projection Chamber (TPC), the Time-Of-Flight detector (TOF) and the V0 scintillators~\cite{ALICE_Det}. The data were collected using a minimum bias trigger, requiring a hit in either one of the V0 scintillators or the Silicon Pixel Detector (SPD) in coincidence with the arrival of proton bunches from both directions. Background contamination was removed offline both using V0 timing information and exploiting the correlation between pixel hits and SPD track segments. Only events with a single primary vertex within $|z| < 10~\textrm{cm}$ were analysed. Furthermore, at least one charged particle with $p_{\textrm{\tiny T}}>0$ in the pseudorapidity region $|\eta|<1$ (i.e.\ INEL > 0 event class) was required. It is to be noted that to avoid auto-correlation biases, the event activity was estimated by the measured signal in the V0 detectors, positioned at $2.8<\eta<5.1$ and $-3.7<\eta<-1.7$, while the charged particle multiplicity density and the identified particles production are measured at mid-rapidity ($|\eta|<0.5$).

\section{Results}
\subsection{Transverse momenutm distributions}

\begin{figure}[t]
  \begin{center}
    \includegraphics[width=.56\textwidth]{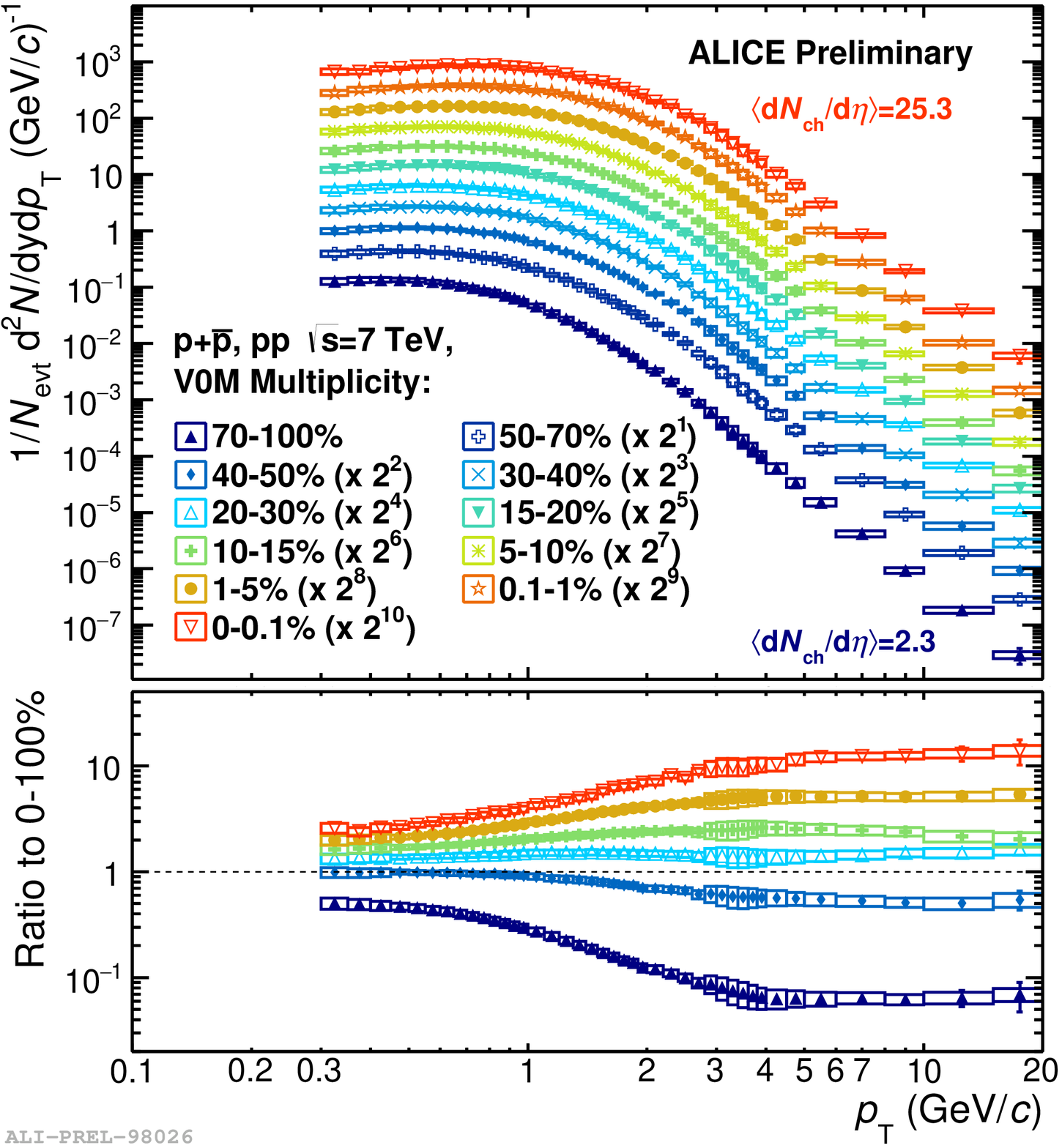}  
    \includegraphics[width=.42\textwidth]{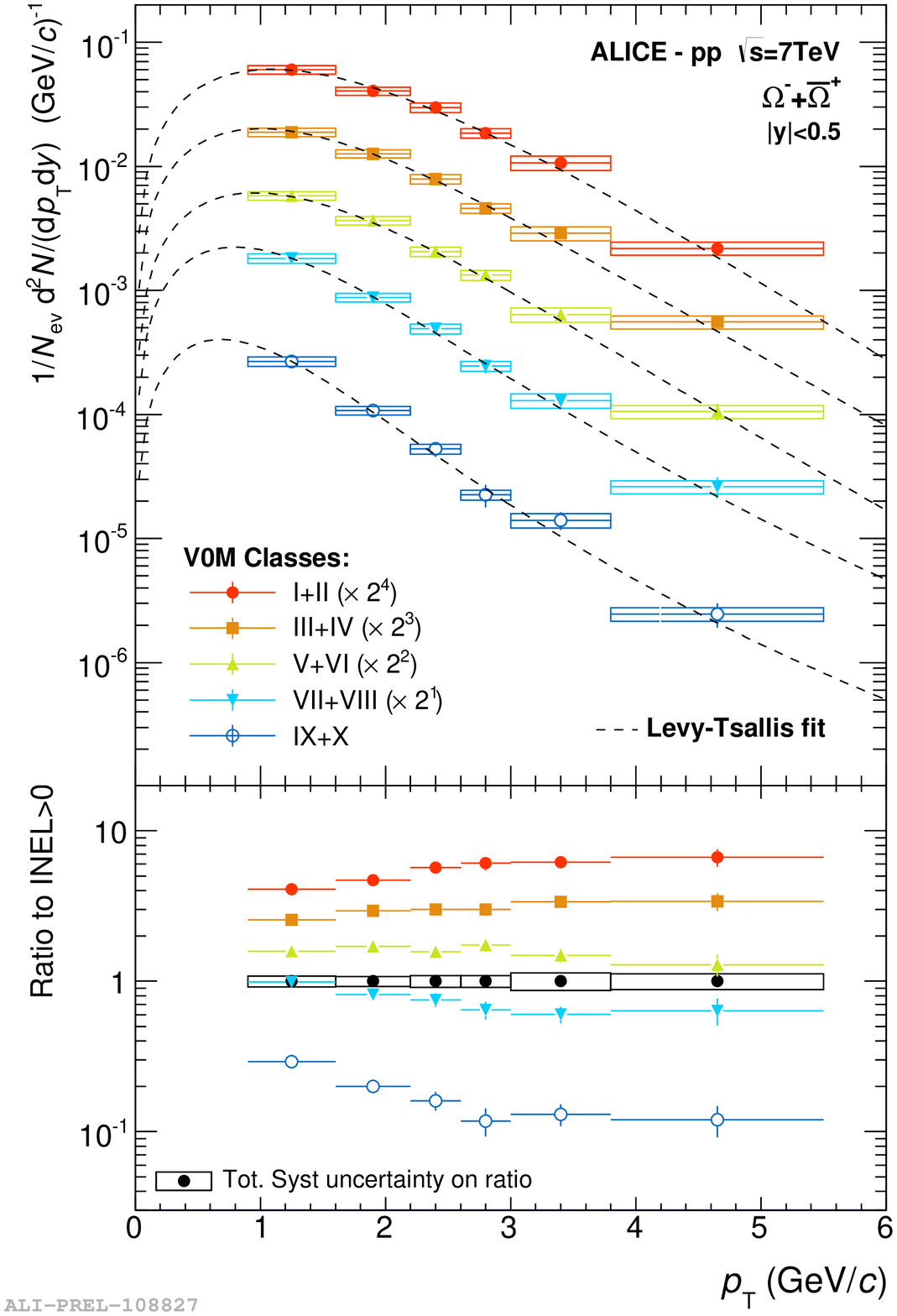}
  \end{center}
  \vspace{-20pt}
  \caption{\pT-differential spectra measured at $|y|<0.5$~as a function of the charged particle multiplicity density in \ppS pp collisions for $\textrm{p}+\bar{\textrm{p}}$  and $\Omega^{-}+\bar{\Omega}^{+}$ (top panels) and their ratio to inclusive spectra (bottom panels).}
    \label{fig:spectra}
\end{figure}

A comprehensive set of light-flavoured particle \pT-differential spectra at midrapidity as a function of charged particle multiplicity density \dNdeta in \ppS pp collisions has been measured, two of them (p and $\Omega$) are shown in \fig{spectra}. The bottom panels report the \pT-differential ratios of spectra in \dNdeta bins to the minimum bias spectrum. 

Transverse momentum spectra exhibit hardening with increasing charged particle multiplicity density. The observed boost in \pT is more pronounced for heavier hadrons, a trend reminescent of that in p-Pb and Pb-Pb collisions~\cite{pPb_PbPb_pOverpi,PbPb_Centrality}. On the other hand, at $p_{\textrm{\tiny T}} \gtrsim 3-5~\textrm{GeV}/c$ ratios to multiplicity-integrated spectra flatten out, hinting of a scaling behaviour in the production of hard particles. Further theoretical developments are needed in order to identify the relevant scaling variable.
\begin{figure}
  \begin{center}
    \includegraphics[height=.3\textheight]{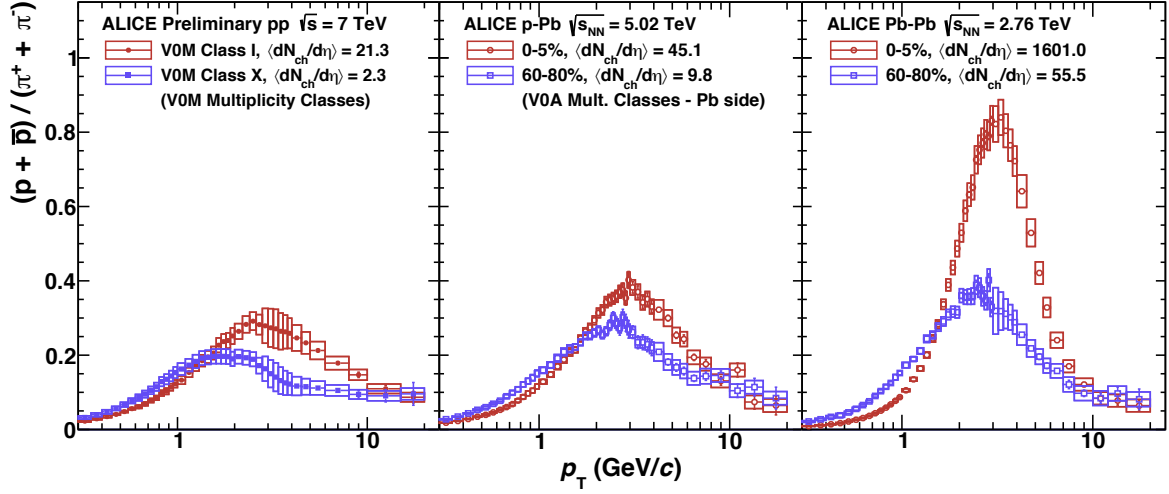}
    \caption{\pT-differential $\textrm{p}/\pi$ ratios in pp at \ppS (left), p-Pb at \S{5.02} (middle) and Pb-Pb at \S{2.76} (right) collisions in low- and high-multiplicity density bins for each collision system.~\cite{pPb_PbPb_pOverpi}}
    \label{fig:poverpi}
  \end{center}
\end{figure}

Transverse momentum spectra evolution with multiplicity is typically studied to investigate recombination and radial flow effects by comparing baryon-to-meson ratios at high and low multiplicities. In pp and p-Pb collisions, a depletion is observed in the $\textrm{p}/\pi$ ratio for $\pT < 0.7~\textrm{GeV}/c$ (\fig{poverpi}), followed by an enhancement at intermediate \pT (above 3 $\textrm{GeV}/c$), a feature first observed in nucleus-nucleus collisions and usually associated with either coalescence or radial flow~\cite{Coalescence_1, Coalescence_2, Flow}. Quantitatively, the observed effects are more pronounced in p-Pb and Pb-Pb collisions, but  \dNdeta are also of different magnitudes.

The mass ordering observed in Pb-Pb is commonly associated with a hydrodynamical evolution of the system, a prerequisite of which is local thermal (kinematic) equilibrium; thermal equilibrium implies a chemically equilibrated system, which can be studied by looking at the chemical composition of particle production, i.e.\ \pT-integrated yields.

\subsection{Integrated yields}

Particle yields were calculated by integrating \pT-differential spectra in the measured range and extrapolating to zero using the L\'evy-Tsallis parametrization~\cite{LeviTsallis}. Systematic uncertainties were obtained by replacing the functional form with Boltzmann, $m_{\textrm{\tiny T}}$-exponential, \pT-exponential, Fermi-Dirac and Bose-Einstein functions.

\begin{figure}
  \begin{center}
    \includegraphics[width=0.52\textwidth]{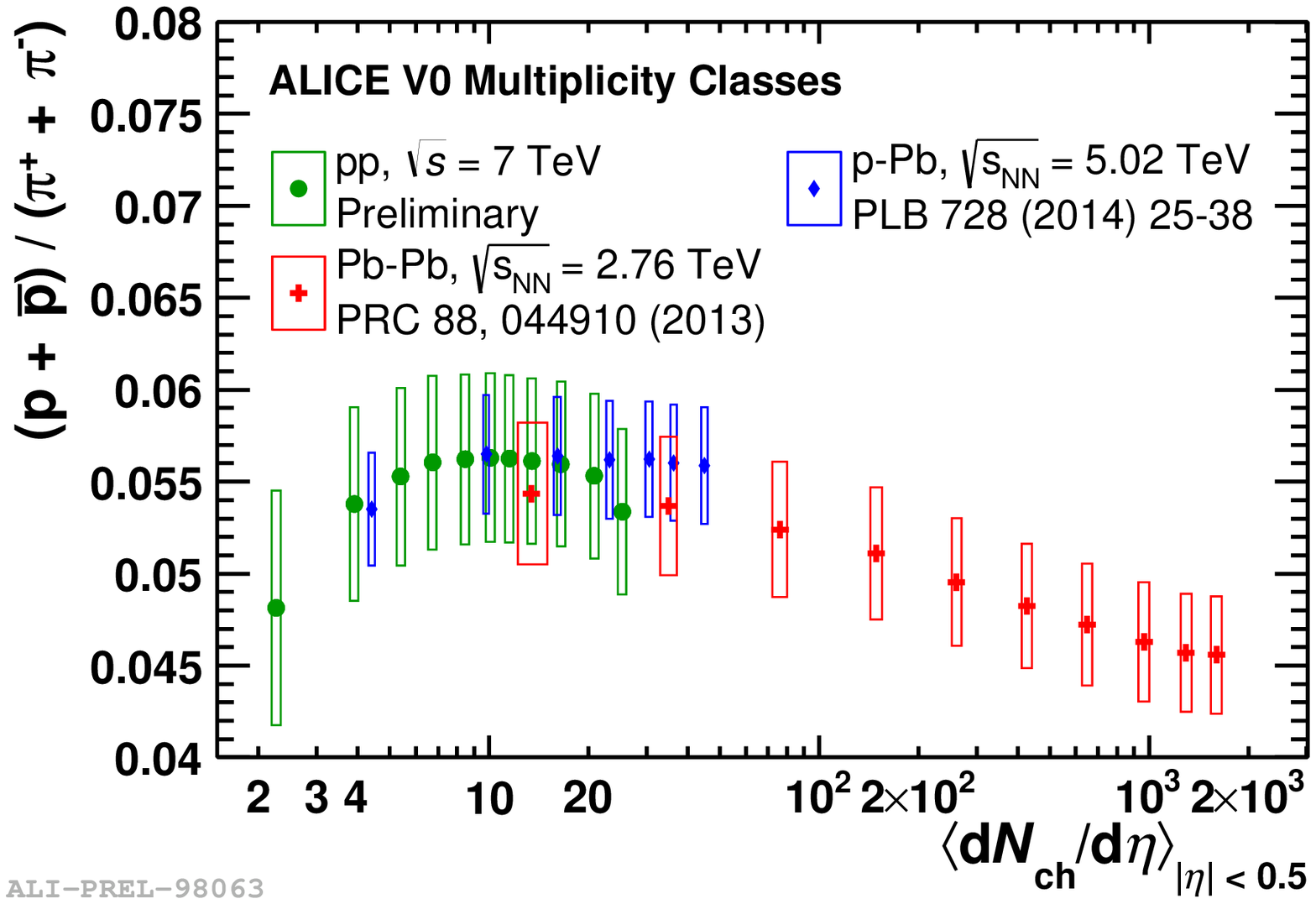}
    \includegraphics[width=0.44\textwidth]{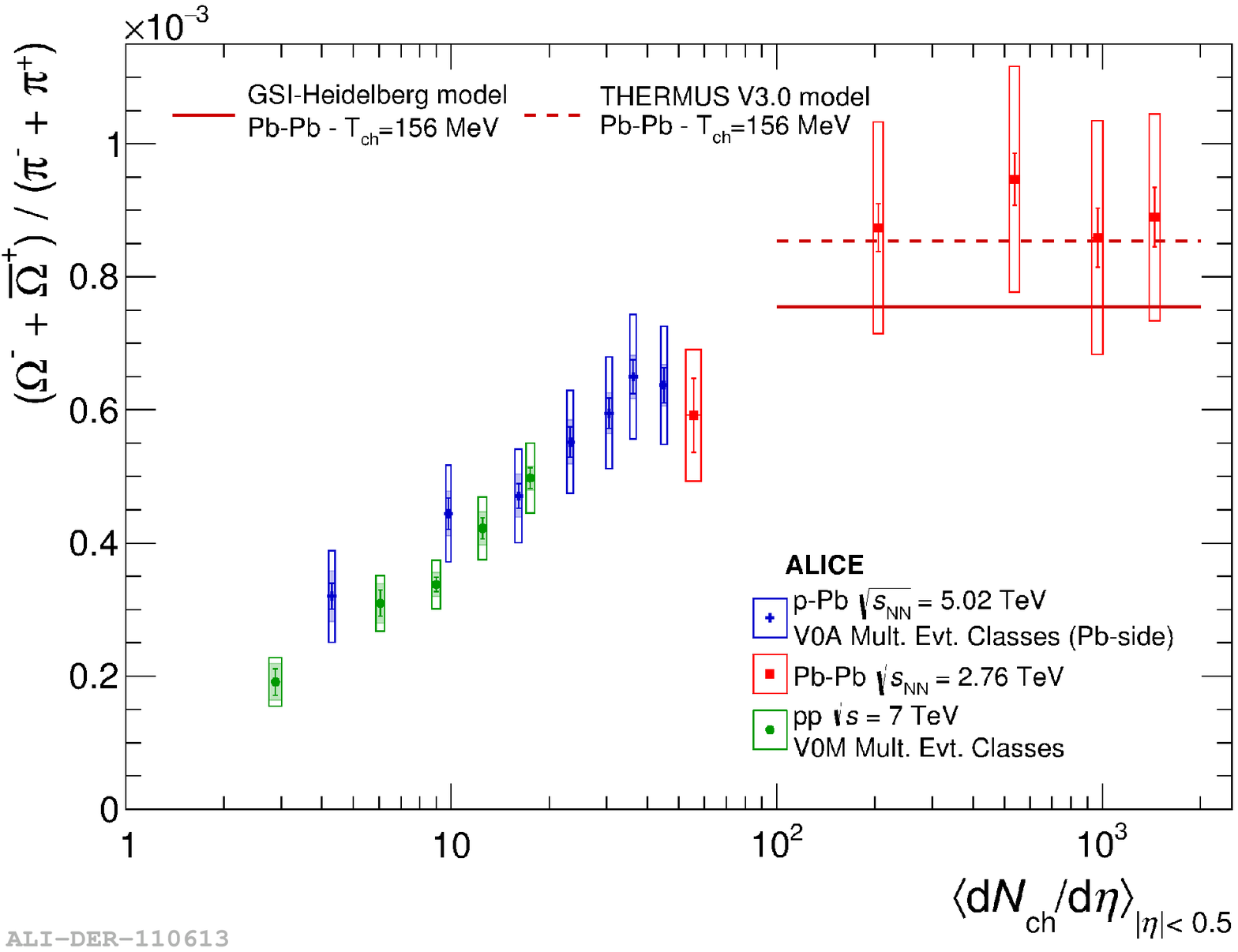}
  \end{center}
  \caption{Integrated $\textrm{p}/\pi$ (left) and $\Omega/\pi$ (right) ratios as a function of charged particle multiplicity.}
  \label{fig:boverpi}
\end{figure}

Particle yield ratios $\textrm{p}/\pi$ and $\Omega/\pi$ as a function of charged particle multiplicity density are shown in \fig{boverpi}. Remarkably, there is a smooth transition from pp to p-Pb and then to Pb-Pb at similar \dNdeta. While $\textrm{p}/\pi$ is essentially constant in pp collisions, $\Omega/\pi$ exhibits a significant increase with event activity, a trend that is also observed when studying other particle species having a smaller strangeness quantum number, albeit weaker in those cases.

\begin{figure}[!h]
        \begin{center}
                \includegraphics[width=0.45\textwidth]{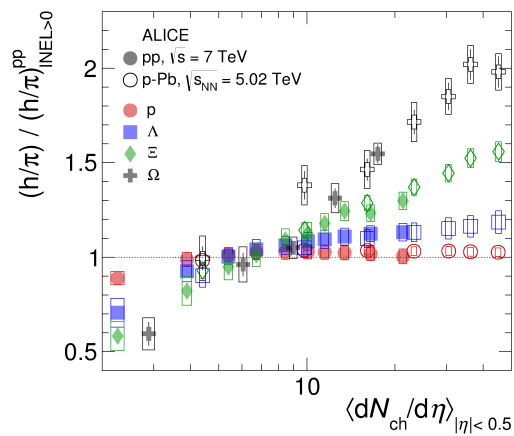}
        \end{center}
        \caption{\pT-integrated baryon-to-pion ratios in pp and p-Pb as a function of charged particle multiplicity density normalized to inclusive baryon-to-pion ratios for pp and p-Pb collisions~\cite{DoubleRatio}.}
        \label{fig:doubleratio}
\end{figure}

A summary of the baryon-to-pion ratios normalized to multiplicity-integrated ratios is shown in \fig{doubleratio}~\cite{DoubleRatio} for pp and p-Pb collisions. The $\textrm{p}/\pi$ double-ratios are consistent with unity in all but the lowest \dNdeta bins. On the contrary, (multi-)strange double-ratios exhibit an increase with the event activity, with a larger slope for baryons with higher strangeness content. This indicates a strangeness-related enhancement of particle production versus multiplicity and not an enhancement related to baryon number.
\clearpage

\subsection{Comparison to Thermal Model calculations}

\begin{figure}
        \begin{center}
        \includegraphics[width=0.6\textwidth]{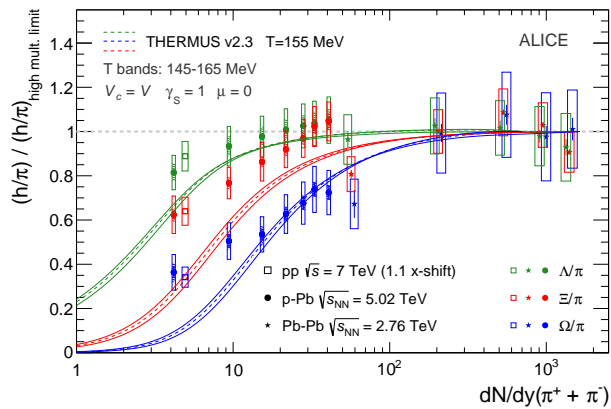}
        \end{center}
        \caption{Strange baryon-to-pion ratio as a function of charged pion multiplicity in comparison to THERMUS v2.3 predictions. Both data and model predictions are normalized to the corresponding ratios at highest multiplicities available.}
        \label{fig:thermuscomp}
\end{figure}

Abundances of hadron production observed in heavy-ion collisions can be qualitatively described by statistical hadronization models, such as THERMUS~\cite{THERMUS}. Typically, statistical calculations for heavy-ion collisions are carried out in the grand-canonical ensemble, which is only applicable provided the system is large enough for each conserved quantum number, e.g.\ flavor.  An attempt to extend the applicability of one of the models to smaller systems, namely p-Pb and pp, is presented in \fig{thermuscomp}~\cite{THERMUSFit}. Model predictions were calculated varying the fireball volume. Three different temperatures in range of $146-166~\textrm{MeV}$ were used for systematics, but each time the model would describe the particles with the same T. Note that pp results shown are multiplicity-integrated.

All baryon-to-pion ratios in consideration approach the Grand Canonical limit following a trend predicted by THERMUS model, while the existing tunes of QCD-inspired Monte Carlo generators show little to no multiplicity dependence~\cite{MCGen}.

\section{Summary}

The ALICE collaboration has reported the production of a comprehensive set of light flavoured particle spectra as a function of charged particle multiplicity density in \ppS pp collisions. Transverse momentum distributions exhibit a hardening with increasing event multiplicity which is more pronounced for heavier particles. These trends are reminiscent of those observed in p-Pb and Pb-Pb systems.

Integrated hadron-to-pion ratios are consistent between pp, p-Pb and Pb-Pb at similar \dNdeta, indicating that the chemical composition of the events seems to be driven by the charged particle multiplicity rather than the initial state of the colliding system. Moreover, the proton-to-pion ratio shows no significant evolution with \dNdeta in pp collisions, while strange baryon-to-pion ratios exhibit a strangeness-related enhancement with multiplicity density. The same trend is qualitatively reproduced by the THERMUS model in p-Pb, as opposed to existing MC generator tunes.

\end{document}